\documentclass[preprint,aps,pre]{revtex4}
\usepackage{graphicx}
\usepackage{natbib}
\usepackage{color}
\usepackage{amsmath}
\usepackage{epstopdf}

\newcommand\Ca{\mbox{\textit{Ca}}}

\newsavebox{\astrutbox}
\sbox{\astrutbox}{\rule[-5pt]{0pt}{20pt}}

\newcommand{\gdot}[0]{{\dot{\gamma}}}
\newcommand{\vtr}[1] {\mathbf{#1}}

\begin{document}

\title{Moving contact line dynamics: from diffuse to sharp interfaces}

\author{H. Kusumaatmaja}
\email{halim.kusumaatmaja@durham.ac.uk}
\author{E. J. Hemingway}
\email{e.j.hemingway@durham.ac.uk}
\author{S. M. Fielding}
\email{suzanne.fielding@durham.ac.uk}
\affiliation{Department of Physics, Durham University, South Road, Durham DH1 3LE, UK}

\begin{abstract}

  We reconcile two scaling laws that have been proposed in the
  literature for the slip length associated with a moving contact line
  in diffuse interface models, by demonstrating each to apply in a
  different regime of the ratio of the microscopic interfacial width
  $l$ and the macroscopic diffusive length $l_D= (M\eta)^{1/2}$, where
  $\eta$ is the fluid viscosity and $M$ the mobility governing
  intermolecular diffusion.  For small $l_D/l$ we find a diffuse
  interface regime in which the slip length scales as $\xi \sim
  (l_Dl)^{1/2}$.  For larger $l_D/l>1$ we find a sharp interface
  regime in which the slip length depends only on the diffusive
  length, $\xi \sim l_D \sim (M\eta)^{1/2}$, and therefore only on the
  macroscopic variables $\eta$ and $M$, independent of the microscopic
  interfacial width $l$.  We also give evidence that modifying the
  microscopic interfacial terms in the model's free energy functional
  appears to affect the value of the slip length only in the diffuse
  interface regime, consistent with the slip length depending only on
  macroscopic variables in the sharp interface regime.  Finally, we
  demonstrate the dependence of the dynamic contact angle on the
  capillary number to be in excellent agreement with the theoretical
  prediction of \cite{Cox1986}, provided we allow the slip length to
  be rescaled by a dimensionless prefactor. This prefactor appears to
  converge to unity in the sharp interface limit, but is smaller in
  the diffuse interface limit. The excellent agreement of results
  obtained using three independent numerical methods, across several
  decades of the relevant dimensionless variables, demonstrates our
  findings to be free of numerical artifacts.

\end{abstract}

\maketitle


\section{Introduction}

Understanding the way in which a contact line (the line along which an
interface separating two immiscible fluids intersects a solid boundary
wall) moves under an imposed flow is a problem of fundamental
importance in wetting dynamics. It underpins a wide range of phenomena
in nature, for example in the functional adaptation of many biological
systems \citep{Parker2001,Zheng2010,Hu2003}, as well as in
technological applications: from oil recovery \citep{Morrow1990} to
inkjet printing and microfluidics \citep{Tabeling_2010}. From these
examples, it has become clear that experimental progress needs to be
accompanied by the development of physical models and computational
methods that can accurately capture wetting dynamics, even in complex
geometries.

Any purely macroscopic treatment -- in which the two fluids are
considered to be separated by a strictly sharp interface, and subject
to a strictly no-slip condition on the fluid-velocity at the solid
wall -- has long been recognised to predict a non-integrable stress
singularity for any non-zero velocity of the contact line
~\citep{Huh1971, deGennes1985}.  This so-called moving contact line
singularity is clearly at odds with the commonplace experience that
the contact line does, in fact, move.

To regularise this singularity, the macroscopic picture just described
must be supplemented by additional physics that intervenes on shorter,
microscopic lengthscales.  One approach is to relax the no-slip
boundary condition on the fluid velocity, by inserting a slip boundary
condition in a region of microscopic size $\xi$ in the vicinity of the
contact line.  This can be done by hand \citep{Zhou_1990,Spelt_2005},
for example by imposing a slip velocity $v_{\rm s}=U\exp(-|x|/\xi)$,
where $x$ is the distance along the wall away from the contact line,
or by choosing $v_{\rm s}=\xi\sigma/\eta$, where $\sigma$ and $\eta$
are the wall stress and fluid viscosity respectively. In any
computational study, slip can also arise indirectly as a numerical
artifact \citep{Renardy2001}.

Another approach is to keep the no-slip boundary condition and instead
to recognise that any interface between two fluids can never be
perfectly sharp, but in practice must have some non-zero microscopic
diffuse width $l$. Accordingly, the use of diffusive interface models
\citep{Anderson1998,Jacqmin2000} has gained increasing popularity in
recent years.  A diffuse interface model of liquid-gas coexistence
then accommodates contact line motion by a condensation-evaporation
mechanism which, in some small region within close proximity of the
contact line, slowly transfers matter from one side of the interface
to the other
\citep{Seppecher_1996,Briant2004a,Pooley2009,Diotallevi_2009}.
Similarly, a diffuse interface model of coexisting immiscible binary
fluids accommodates contact line motion via a slow intermolecular
diffusion of the two fluids across the interface between them, again
acting in a small region in the vicinity of the contact line
\citep{Jacqmin2000,Briant2004b,Pooley2009}.


Diffuse interface models \citep{Anderson1998,Jacqmin2000} are also
convenient in computational practice. An order parameter (``phase
field'') is introduced to distinguish one fluid from another, with the
length scale that characterises the variation of this field across the
interface between the fluids prescribing the interfacial width, $l$.
An advection-diffusion equation (the Cahn-Hilliard equation
supplemented by an advective term) then determines the evolution of
this order parameter, and hence of the interface position. This is
solved in tandem with the Navier-Stokes equation for the fluid
velocity field.  A key advantage of this approach is that all
computational grid points are treated on an equal footing, without any
need to explicitly track the position of the interface.  In this way,
even complex flow geometries can be simulated conveniently.

It is important to emphasize, as first elucidated by ~\cite{Cox1986},
that although the physical origins of contact line motion may differ
according to the detailed microscopic physics invoked (wall slip,
intermolecular diffusion across the interface, {\it etc.}), the
macroscopic hydrodynamics far from the contact line nonetheless
converges to a universal solution that is informed by the microscopics
only in the sense of being rescaled by a single parameter, the ``slip
length'' $\xi$, that emerges from this underlying microscopic physics.
 An important problem within any model of the microscopics is
therefore to determine this emergent slip length $\xi$: not only
because it is a key variable that determines the macroscopic wetting
and fluid dynamics, but also because it sets a scale to which other
length scales (surface heterogeneities, droplet size, {\it etc.})
must be compared.

With that motivation in mind, the primary aim of this work is to
determine the scaling of the slip length within a diffuse interface
model of immiscible binary fluids.  We assume the fluids to have
matched viscosity $\eta$, and denote by $M$ the mobility parameter
characterising the rate of intermolecular diffusion in the
Cahn-Hilliard equation.  In the existing literature, (at least) three
apparently contradictory scalings have been proposed.  Several authors
(e.g.  \cite{Jacqmin2000}, \cite{Yue2010}) suggest that $\xi \sim l_D$
where $l_D=(M\eta)^{1/2}$ is the characteristic lengthscale below
which intermolecular diffusion dominates advection and above which the
opposite is true. For convenience we call this the diffusion length in
what follows, and we emphasise that it is determined only by the
macroscopic quantities $M$ and $\eta$. In contrast, \cite{Briant2004b}
suggest a different scaling, $\xi\sim \sqrt{l_Dl}$, which is the
geometric mean of the macroscopic diffusion length and the microscopic
interface width. And while these two proposed scalings do not have any
dependence on the imposed flow velocity, \cite{Chen2000} in contrast
suggested that the slip length does depend on the velocity as $\xi
\propto V_0^{-1/2}$.

In view of these apparently contradictory predictions, an important
contribution of this work is to provide a coherent picture that fully
reconciles two of these existing predictions, by showing each to apply
in a different regime of parameter space, and that carefully appraises
the third. This is achieved by extensive numerical studies performed
across unprecedently wide ranges of the two relevant control
parameters. Specifically, we explore four decades of the dimensionless
ratio $M\eta/l^2=l_D^2/l^2$ of the (squared) macroscopic diffusion
length to the (squared) microscopic interface width; and three decades
of the capillary number $\Ca$ (defined below), which adimensionalises
the imposed flow velocity $V_0$ in terms of an intrinsic interfacial
velocity scale.  Moreover we use three different, independently
coded numerical methods, and show their results to be in excellent
quantitative agreement across all decades.

For small values of the capillary number $\Ca$, corresponding to low
imposed flow velocities, we demonstrate that the slip length converges
to a well-defined value that is independent of the velocity $V_0$. Any
dependence on the velocity is only observed for $\Ca > 0.01$, and we
comment on the extent to which our results are consistent with the
prediction $\xi\propto V_0^{-1/2}$ of \cite{Chen2000} in this fast
flowing regime.

The remainder of the paper focuses on the slow flow regime, $\Ca <
0.01$, in which the slip length $\xi$ is independent of the flow rate
$V_0$.  Here we show that, in the limit of large $l_D/l$, the slip
length scales as $\xi \propto l_D$, informed only by the macroscopic
lengthscale $l_D$. This agrees with the original prediction of
\cite{Jacqmin2000}, \cite{Yue2010} and corresponds to the sharp
interface limit of the diffuse interface model. In this limit the
microscopic length $l$ essentially drops out of the problem, apart
from providing a ``singular perturbation'' to regularise the
contact line singularity.  In contrast, for small $l_D/l$ we find that
the slip length scales as $\xi \propto (l_Dl)^{1/2}$, as proposed by
\cite{Briant2004b}. This corresponds to the diffuse interface limit in
which the emergent slip length does depend on the underlying
microscopic length $l$.  In this way we reconcile these two previously
apparently contradictory scalings, by showing each to apply in a
different limiting regime of $l_D/l$.  The crossover between the two
is furthermore consistent with the onset of the sharp interface regime
discussed by \cite{Yue2010}.

The fact that the slip length $\xi$ is in general different from the
interface width $l$, and indeed greatly exceeds it for large $l_D/l$,
has a clear practical consequence for any simulation: the resolution
of the full wetting dynamics appears only on a lengthscale
corresponding to the larger of the interface width and the slip
length.  This, in turn, limits the region of parameter space that
can be considered reliable in any numerical study.

We provide additional evidence for the distinction between the diffuse
interface limit $l_D/l \ll 1$, in which the microscopic physics
informs $\xi$, and the sharp interface limit $l_D/l\gg 1$, in which it
does not, by performing further simulations in which the we modify the
interfacial (microscopic) contribution to the underlying free energy
functional of the binary fluid.  We show that the slip length depends
on this modification only in the diffuse interface regime. In the
sharp interface regime of large $l_D/l$, the slip length continues to
depend only on the macroscopic dynamical variables, $\eta$ and $M$,
free of this modification to the microscopic interfacial term.

Finally, we test the validity of Cox's formula \citep{Cox1986} for the
dependence of the dynamic contact angle on the Capillary number in the
different slip length regimes. Our numerical results are in good
agreement with Cox's analytical result if we allow the slip length to
be rescaled by a dimensionless parameter. Moreover this parameter
appears, suggestively, to converge to unity in the sharp interface
limit, but is smaller in the diffuse interface limit.

This paper is organized as follows. In the next section we describe
the models, methods and setups we employ to compute the slip lengths
and dynamic contact angles. We present our results in section
\ref{sec:results}. Finally, we summarise our key findings and discuss
avenues for further work in section \ref{sec:conclusion}.

\section{Model, geometry and boundary conditions}
\label{sec:model}

In this section we specify the model that we shall use throughout the
paper, starting with the thermodynamics in Secs.~\ref{sec:thermo}
and~\ref{sec:curvatureDef} then the dynamical equations of motion in
Sec.~\ref{sec:dynamics}. We then specify the flow geometry and
boundary conditions in Sec.~\ref{sec:geometry}.

\subsection{Thermodynamics}
\label{sec:thermo}

We consider a binary mixture of two mutually phobic fluids, $A$ and
$B$, and denote the volume fraction of fluid $A$ by the continuum
phase field $\phi(\vtr{r},t)$.  The volume fraction of fluid $B$ is
then simply 1-$\phi$ by mass conservation and need not be considered
separately.  We consider a Landau free energy~\citep{Bray1994}
\begin{eqnarray}
  \Psi = \int_V \psi_b \,\, dV = \int_V \left[ \frac{G}{4} (\phi^2-1)^2 + \frac{Gl^2}{2} (\nabla \phi)^2 \right] dV,
  \label{eq:freeen}
\end{eqnarray}
which allows a coexistence of two bulk phases: an A-rich phase with
$\phi_A = 1$ and a B-rich phase with $\phi_B = -1$. The bulk constant
$G$ has dimensions of energy per unit volume (or equivalently of force
per unit area, and so modulus). If the two fluids have different
affinities to the solid surface, a surface (wetting) contribution to
the free energy can be added~\citep{Cahn, Pooley2008, Pooley2009} to
the right hand side of Eq. \eqref{eq:freeen}. Throughout this work we
assume neutral wetting conditions, for which no such contribution is
needed.

The chemical potential $\mu$ follows as a functional derivative of the
free energy density $\psi_b$ with respect to $\phi$, giving
\begin{equation}
  \mu = - G \phi + G \phi^3 - Gl^2 \, \nabla^2 \phi.
  \label{eq:chempot}
\end{equation}
In equilibrium the chemical potential $\mu = 0$. Assuming a flat
interface of infinite extent in the $y-z$ plane, with the interfacial
normal in the $x$ direction and the interface located at $x = 0$, we
then obtain an interfacial solution of the form
\begin{equation}
  \phi = \tanh \left(\frac{x}{\sqrt{2}l}\right),
  \label{eq:Sol}
\end{equation}
with a homogeneous B-rich phase for $x \ll -l$ and homogeneous A-rich
phase for $x \gg l$. The interfacial constant $l$ specifies the
characteristic length scale over which $\phi$ varies in between these
two phases, and so corresponds to the interfacial width.  The surface
tension associated with this interface is given by
\begin{equation}
  \sigma = \frac{2 \sqrt{2} Gl}{3}.
\end{equation}

\subsection{Generalisation of the Landau Theory: Introduction of a
  curvature term }
\label{sec:curvatureDef}

So far, we have specified the Landau free energy in the form most
commonly used in the literature. It will also be instructive in what
follows to consider the extent to which our results for the slip
length do or don't depend on the microscopic details of the model
used. Accordingly, we now generalise the free energy slightly to give
\begin{equation}
  \Psi' = \int_V \psi'_b \,\, dV = \int_V \left[ \frac{G}{4} (\phi^2-1)^2 + \frac{Gl^2}{2} (\nabla \phi)^2 + \alpha\frac{Gl^4}{4} (\nabla^2 \phi)^2 \right] dV.
  \label{eq:fecurvature}
\end{equation}
Compared to the original free energy defined above, this has an
additional interfacial curvature term of amplitude set by $\alpha$.
The mapping between this form of Landau free energy and the
(Helfrich) continuum elastic energy can be found in e.g.
\cite{Gompper1991}. The associated chemical potential is
\begin{equation}
  \mu = G\left(-  \phi + \phi^3 - l^2 \, \nabla^2 \phi - \alpha l^4 \, \nabla^4 \phi \right).
  \label{eq:chempotmod}
\end{equation}
It is important to note that the bulk terms are unaltered, with the
modification affecting only interfacial gradient terms involving
powers of $l\,\nabla$.  A careful comparison of the original model,
for which $\alpha = 0$, with this generalised model, for which $\alpha
> 0$, will allow us to demonstrate that the slip length is independent
of the microscopic details, as specified by $\alpha$, in the sharp
interface regime $l_D/l\gg 1$.  In contrast in the diffuse interface
regime $l_D/l \ll 1$ we find that the slip length does depend on the
microscopics, via $\alpha$.

\subsection{Equations of Motion}
\label{sec:dynamics}

The dynamics of the order parameter field is specified by the
Cahn-Hilliard equation \citep[see e.g.~][]{Bray1994} generalised to
include an advective term:
\begin{equation}
  \left(\partial_t + \vtr{v}.\nabla\right) \phi = M \nabla^2 \mu.
  \label{eq:cahnhilliard}
\end{equation}
Here $M$ is the molecular mobility, which we assume constant.  The
fluid velocity and pressure fields, $\vtr{v}(\vtr{r},t)$ and
$p(\vtr{r},t)$, obey the continuity and Navier-Stokes equations
\begin{eqnarray}
  & \partial_t \rho + \vtr{v}.\nabla \rho = -\rho \nabla.\vtr{v}, \label{eqn:comp}\\
  & \rho \left(\partial_t + \vtr{v}.\nabla\right) \vtr{v} = \eta \nabla^2 \vtr{v} - \nabla p - \phi \nabla \mu.
  \label{eq:navierstokes}
\end{eqnarray}
We denote by $\rho$ and $\eta$ the fluid density and viscosity respectively,
assuming throughout that the two fluids are perfectly matched in both
density and viscosity. In the Navier-Stokes equation, gradients in the
chemical potential contribute an additional forcing term to the fluid
motion, $-\phi \nabla \mu$ \citep{Jacqmin1999}.

Diffuse interface models are widely used in the computational
fluid dynamics community and they have been implemented using various approaches.
To ensure our results are free from numerical artefacts, here
we have used three different numerical methods:
(MI) the lattice Boltzmann method, (MII) spectral method, and (MIII)
immersed boundary method to solve the equations of motion.
They are detailed in Sec.~\ref{sec:numerics} below.
For MII and MIII, we have assumed incompressible flow and taken the inertialess
limit of zero Reynolds number Stokes flow. This corresponds to
setting to zero the
terms on the left-hand-sides of Eqn.~\ref{eqn:comp}
(incompressibility) and Eq.  \ref{eq:navierstokes} (zero inertia). The
Lattice-Boltzmann method (MI) intrinsically requires a small but
non-zero inertia and compressibility, though our numerical
results confirm the effects of this difference to be negligible for
the problem considered here.

\subsection{Geometry, initialisation and boundary conditions}
\label{sec:geometry}


\begin{figure}
  \centering
  \includegraphics[width=\textwidth]{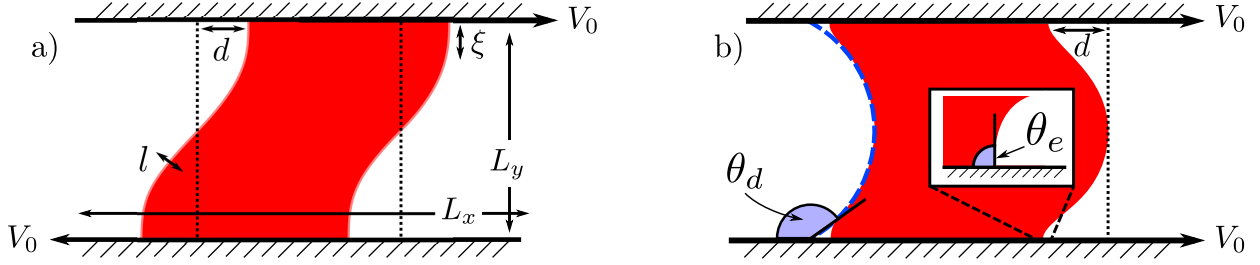}
  \caption{Schematic illustration of typical
    steady-states in our channel geometry. Marked are the key
    lengthscales: slip-length $\xi$, interface displacement $d$,
    channel dimensions $L_x, L_y$, and diffuse interface width $l$. In
    panel a) a shear flow is imposed by moving the plates with equal
    and opposite velocities $\pm V_0$. In panel b) a Poiseuille flow
    is imposed by applying a pressure drop along the length of the
    channel, sketched here in the frame in which the fluid bridge is
    stationary. The definitions for the dynamic ($\theta_d$,
    macroscopic) and equilibrium ($\theta_e$, microscopic) contact
    angles are shown.}
  \label{fig:schematic}
\end{figure}

We consider flow between infinite flat parallel plates a distance
$L_y$ apart, with plate normals in the $y$ direction. In the flow
direction $x$ the cell is taken to have length $L_x = 2 L_y$, with
periodic boundary conditions on both $\phi$ and $\vtr{v}$.  All
quantities are assumed invariant in the $z$ direction.  The phase
field is initialised in an equilibrium state with a bridge of A-rich
phase, in which $\phi = + 1$, separated by two vertical diffuse
interfaces of width $\ell$ at $x = L_x/4$ and $ 3L_x/4$ from B-rich
phases on either side, where $\phi = -1$. The equilibrium contact
angle is taken to be $\theta_e = 90^\circ$, corresponding to neutral
wetting conditions.  Throughout we define the position of the
interface itself by the locus $\phi = 0$.

The fluid is taken to be initially at rest with $\vtr{v}=0$
everywhere. Starting from this initial condition, we then
implement one of two common flow protocols: boundary-driven planar
Couette flow and pressure-driven planar Poiseuille flow. In the first
of these, a constant shear-rate $\gdot$ is applied by moving the top
and bottom plates at velocities $V_0 = \pm \frac{1}{2}\gdot L_y$. This
deforms the phase field and for small $\Ca \equiv \eta V_0 / \sigma
\lesssim 0.1$ a steady state is reached in which the interface has
displaced a distance $d \propto \pm\Ca$ at the top and bottom
walls, as shown in Fig. \ref{fig:schematic}a. The deformed bridge
is then stationary, with the contact lines moving at a velocity
$\mp V_0$ relative to the top and bottom walls.

In the Poiseuille protocol the flow is driven by an imposed pressure
drop $\Delta P$ along the length of the channel. A steady state then
develops in which the bridge migrates along the channel at a constant
speed, with the contact lines moving along each wall with a measured
velocity $V_0$. In the reference frame of the contact line, therefore,
both walls move with a velocity $-V_0$. By fitting the interface in
the central region of the channel (between $y = \frac{1}{4} \to
\frac{3}{4}$) to a circle, see Fig.~\ref{fig:schematic}b, we define the
dynamic contact angle $\theta_d$ as the angle of intersection this
circle makes with the wall. The dynamic contact angle is in general
different from the equilibrium contact angle.

At the plates we assume boundary conditions of no-slip and
no-permeation for the fluid velocity.  For the phase field, we
implement $\partial_y\mu|_{y=0,L_y}=0$ in method MIII.  In method MII
it is instead more convenient to set $\partial_y \phi|_{y = 0, L_y} =
\partial^3_y \phi|_{y = 0, L_y} = \partial^5_y \phi|_{y = 0, L_y} = 0$
(where the last equality need only be imposed in the case
$\alpha \neq 0$).  Note that while this condition automatically
ensures zero gradient of the chemical potential, it is actually a
stronger condition than that in demanding all the relevant odd
derivatives of $\phi$ to vanish separately.  We do, however, find no
difference in our numerical results between these. For method MI, the
bounce-back boundary conditions assure we have no slip and no
permeation of either fluid across the boundary.

\subsection{Definition of the slip length}
\label{sec:sliplength}

\begin{figure}
  \centering
  \includegraphics[width=0.4\textwidth]{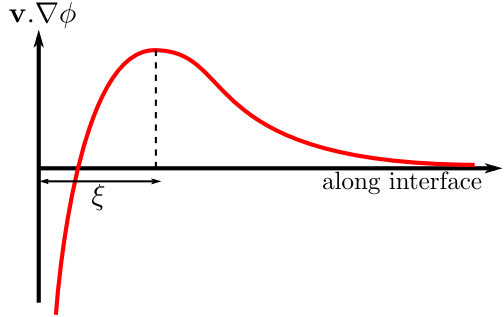}
  \caption{Illustration of our definition of the slip length $\xi$, defined as the position of the first local maximum of $\vtr{v}.\nabla \phi$ as measured along the interface (defined as the locus $\phi = 0$). }
  \label{fig:sliplength}
\end{figure}

Our definition of the slip length uses the fact that the slip mechanism in
the binary fluid model is intermolecular diffusion.
In steady state, the Cahn-Hilliard equation of motion for the order
parameter reads
\begin{equation}
  \vtr{v}.\nabla \phi = M \nabla^2 \mu.
  \label{eq:steadystate}
\end{equation}
In Fig. \ref{fig:sliplength}, we plot the way in which $\vtr{v}.\nabla
\phi$ typically varies along the fluid-fluid interface (taken as the
locus $\phi = 0$, as noted above). We take the slip length
$\xi$ to be given by the distance from the wall to the local maximum.
In measuring this typical distance away from the wall over which
$\vtr{v}.\nabla \phi = M \nabla^2 \mu$ remains appreciable, $\xi$
characterises the lengthscale over which intermolecular diffusion
is important.

Other definitions of the slip lengths have been used in the
literature to characterize the diffusive mechanism. For examples,
\cite{Briant2004b} defined the slip
length as the point at which $M \nabla^2 \mu$ reaches -10\% of the
value at the wall after passing the first maximum, as measured along
the interface;  \cite{Yue2010} suggested the use of the location of the
stagnation point in the flow. As we will show in the results section,
we recover the same scaling laws for the diffuse and
sharp interface limits as in \cite{Briant2004b} and \cite{Yue2010} respectively,
strongly suggesting these definitions are not independent, as expected if there
is a genuine characteristic length scale in the problem.

\subsection{Parameters and dimensionless groups}
\label{sec:parameters}

For the model and flow geometry that we have now defined, we have the
following parameters: the fluid density $\rho$, the fluid viscosity
$\eta$, the modulus $G$ that sets the scale of the free energy
density, the mobility $M$, the diffuse interfacial width $l$, the box
dimensions $L_x$ and $L_y$, and the applied flow velocity $V_0$. (Also
relevant numerically are the spatial mesh size and the timestep;
we have ensured convergence for all the results presented.) As noted above,
the fluid density is set to zero (inertialess flow) for methods MII
and MIII, and small for MI. This leaves $\eta,G, M,l,L_y,L_x$ and
$V_0$ as the relevant parameters. We are then free to choose units of
mass, length and time, thereby reducing by three the number of
parameters that must be explored numerically. Accordingly, we work in
what follows with the following dimensionless groups:
$\sqrt{M\eta}/l$, $l/L_y$, $L_x/L_y$ and the capillary number
$\Ca=\eta V_0/\sigma$. The last of these characterises the speed of
the imposed flow compared to the intrinsic interfacial velocity scale
$\sigma/\eta$ formed by comparing the surface tension to the fluid
viscosity (recall that the surface tension $\sigma \propto Gl$).
Among these we set the aspect ratio of the box $L_x=2 L_y$ throughout.
The ratio $l/L_y$ is always set small, to ensure that the results are
not contaminated by finite box size effects. This leaves
$\sqrt{M\eta}/l$ and $\Ca$ as the two dimensionless groups to be
explored numerically.

\section{Numerical Methods}
\label{sec:numerics}

In this section we describe our three numerical methods. For
convenience we will refer to these throughout the manuscript as MI,
MII and MIII. We emphasize that each was coded independently of the
others, one by each of the authors of the manuscript, thereby providing
confidence that our results are free of algorithmic artefacts.

\subsection{Method I: Lattice Boltzmann Method}

We use a standard free energy lattice Boltzmann method to solve the binary fluid equations of motion. The basic idea behind the lattice Boltzmann algorithm is to associate distribution functions, $\vtr{f}({\bf r}, t)=\{f_i({\bf r}, t)\}$ and $\vtr{g}({\bf r}, t)=\{g_i({\bf r}, t)\}$, discrete in time and space, to a set of velocity directions $\mathbf{e}_i$. Here we use a D2Q9 model, where $\mathbf{e}_i = (0,0), (\pm 1,0), (0,\pm 1), (\pm 1,\pm 1)$. The physical variables are related to the distribution functions by
\begin{equation}
  \rho=\sum_i f_i , \;\;\;\;\;\; \rho v_{\alpha} = \sum_i f_i e_{i\alpha} ,  \;\;\;\;\;\; \phi = \sum_i g_i, \;\;\;\;\;\; \phi v_{\alpha} = \sum_i g_i e_{i\alpha} .
  \label{eq:LBmoments}
\end{equation}

The time evolution equations for the distribution functions can be broken down into two steps. In the collision step, we have used a single relaxation time approach for the order parameter and a multiple relaxation time approach for the fluid density,
\begin{eqnarray}
  &\vtr{f}^\prime({\bf r}, t) = \vtr{f}({\bf r}, t) - \vtr{M^{-1}SM} [\vtr{f}({\bf r}, t)-\vtr{f}^{eq}({\bf r}, t)]\, , \nonumber \\
  &\vtr{g}^\prime({\bf r}, t) = \vtr{g}({\bf r}, t) - \frac{1}{\tau_\phi} \left[\vtr{g}({\bf r}, t) - \vtr{g}^{eq}({\bf r}, t) \right].
\end{eqnarray}
As shown by \cite{Pooley2008}, the multiple relaxation time approach
is significantly more reliable to reduce spurious velocities at the
contact line. $\vtr{f}^{eq}$ and $\vtr{g}^{eq}$ are local equilibrium
distribution functions. The choice of the local equilibrium functions
must recover the correct thermodynamics and hydrodynamics equations of
motion in the continuum limit. The matrix $\vtr{M}$ performs a change
of basis to more physically relevant variables, including the
  density, momentum, and viscous stress tensor. The matrix $\vtr{S}$
  is a diagonal matrix and contains the information about how fast
  each of these physical variables relaxes to equilibrium at every
  time step. Detailed expressions for $\vtr{f}^{eq}$, $\vtr{g}^{eq}$,
  $\vtr{M}$, and $\vtr{S}$ can be found in \cite{Pooley2008}.

The relaxation parameter $\tau$ for the viscous stress tensor in
$\vtr{S}$ can be related to the kinematic shear viscosity
$\nu=\eta/\rho$ via
\begin{equation}
  \nu = (\Delta t(\tau-1/2))/3 \, .
  \label{eq:ShearViscosity}
\end{equation}
Similarly, the relaxation parameter $\tau_{\phi}$ for the order parameter is related to the mobility parameter in the Cahn-Hilliard equation through
\begin{equation}
  M =  \Delta t \Gamma \left(\tau_\phi - \frac{1}{2}\right) .
  \label{eq:Mobility}
\end{equation}
$\Gamma$ is a constant and $\Delta t$ is the simulation time step. In practice, the relaxation times can therefore be tuned to match the desired continuum dynamical variables, $M$ and $\eta$.

In the propagation step, the updated distribution functions are passed on to the neighbouring lattice points,
\begin{eqnarray}
  &\vtr{f}({\bf r} + {\bf e}_i \Delta t , t+\Delta t) = \vtr{f}^\prime({\bf r}, t) \, ,
  \nonumber \\
  &\vtr{g}({\bf r} + {\bf e}_i \Delta t , t+\Delta t) = \vtr{g}^\prime({\bf r}, t) .
  \label{eq:latbolt}
\end{eqnarray}
At the two walls we use a bounce-back rule for both the $\{f_i({\bf
  r}, t)\}$ and $\{g_i({\bf r}, t)\}$ distribution functions. These
ensure boundary conditions of no slip, no permeation, and no
diffusion for the fluid material across the boundary \citep{Ladd2001}.

\subsection{Method II: Spectral Method}

Within this method, at each numerical timestep we separately (a) solve
the hydrodynamic sector of the dynamics to update the fluid velocity
field $\vtr{v}$ at fixed phase field $\phi$, then (b) update the phase
field at fixed fluid velocity field. In part (a) we use a
streamfunction formulation to ensure that the incompressibility
condition is automatically satisfied. After eliminating the pressure
by taking the curl of the generalised Stokes equation
(Eqn.~\ref{eq:navierstokes} with the left hand side set to zero) we
solve for the streamfunction using a Fourier spectral method.

For the phase field dynamics in (b) we use a third order upwind scheme \citep{Pozrikidis2011}
to update the convective term on the
left hand side of Eqn.~\ref{eq:cahnhilliard}. The gradient terms on
the right hand side are solved using a Fourier spectral method using
both sine and cosine modes in the periodic $x$ direction and only
cosine modes in the $y$ direction for consistency with the imposed
boundary conditions $\partial_y \phi =
\partial^3_y \phi = \partial^5_y\phi = 0$ at the walls $y = 0, L_y$.

For the time-stepping we adopt a method originally proposed by
\cite{Eyre1998}, and later studied in depth by
\cite{Guillen-Gonzalez2013}. This splits the free energy term into an
expansive part, $-\nabla^2\phi$, and a contractive part,
$\nabla^2(\phi^3)$, $-\nabla^4\phi$. These are then distributed
between the $n^{th}$ and $(n+1)^{th}$ timesteps as
\begin{equation}
  \frac{\phi^{n+1} - \phi^{n}}{\Delta t} = M G \left[ \left(-l^2 \nabla^4 \phi + 2 \nabla^2 \phi\right)^{n+1} + \left(\nabla^2 \phi^3 - 3 \nabla^2 \phi\right)^n \right],
  \label{eq:contact_eyre_scheme}
\end{equation}
where $\Delta t$ is the timestep. This method permits significantly
larger timesteps than, for example, a Crank-Nicolson
algorithm~\citep{NumericalRecipes} in which
all terms are split equally between the $n^{th}$ and $(n+1)^{th}$
timesteps. When present, for $\alpha>0$, the sixth order gradient term
is treated fully implicitly.

\subsection{Method III: Immersed Boundary Method}

In methods I and II, the walls of the flow cell were included by means
of a simulation box that is closed in the flow gradient direction $y$.
In method III we instead use a biperiodic box that conveniently
enables us to solve both the hydrodynamic sector of the dynamics
(again in the incompressible streamfunction formulation at zero
Reynolds number) and the diffusive part of the concentration dynamics
using fast Fourier transforms in both spatial dimensions.

To incorporate the walls of the flow cell at $y=0,L_y$, we then
include a set of immersed boundary forces using smoothed (Peskin) delta
functions as source terms in the Stokes equation along the desired
location of each wall, as discussed in ~\cite{Peskin2002} and
\cite{Lai2000}. The force required to ensure zero velocity at the
location of each delta function is then evolved as
\begin{equation}
  \partial_t \vtr{F} = -\kappa\left(\vtr{V} - \vtr{V}_0\right),
  \label{eq:contact_IBM}
\end{equation}
where $\vtr{V}_0$ is the prescribed wall velocity. Here $\kappa$ is a
spring constant, and all the results shown below have been converged
to the limit $\kappa \to \infty$ in successive runs.

For the phase field, we assume the boundary conditions $\partial_y\mu=0$
at the plates $y=0,L_y$. This is implemented in an analogous way to the
no-slip boundary condition, by means of adding an extra source term contribution
\begin{equation}
\partial_t\phi = -K \hat{\vtr{n}}.\nabla \mu
\end{equation}
to the equation of motion for the concentration at the location of the
Peskin delta functions, where $\hat{\vtr{n}}$ is a unit vector normal
to the wall.


\section{Results}
\label{sec:results}

We now present our results. We start in Sec.~\ref{sec:standard} by
considering the standard Landau free energy with $\alpha=0$, before in
Sec.~\ref{sec:curvature} appraising the robustness of our findings to
the inclusion of the additional interfacial gradient (curvature) term
in the free energy, setting $\alpha>0$. In Sec~\ref{sec:Cox}, we
compare our numerical results for the dynamic contact angles to the
analytical predictions by \cite{Cox1986}.

\subsection{Standard Landau theory}
\label{sec:standard}

As discussed previously, in the existing literature two apparently
contradictory scaling laws have been proposed for the contact line's
slip length $\xi$.  Several authors, including \cite{Jacqmin2000} and
\cite{Yue2010}, have proposed that the slip length is proportional to
the diffusive lengthscale $l_D$, which describes the length below
which intermolecular diffusion dominates advection and above which the
opposite is true, giving $\xi\propto l_D = (M\eta)^{1/2}$. In
contrast, the lattice Boltzmann simulations and scaling arguments of
\cite{Briant2004b} suggest that the slip length depends not only on
this macroscopic diffusive lengthscale $l_D$, but also on the
microscopic interfacial width $l$, with $\xi \propto (l_Dl)^{1/2}$.

To resolve this apparent discrepancy we have performed extensive
numerical simulations of the moving contact line problem across four
decades of the relevant dimensionless control parameter
$M\eta/l^2=l_D^2/l^2$. Moreover, to ensure that our results are free
of algorithmic artefacts we have used (across the entire range) the
three different, independently designed and coded numerical techniques
just described.  We focus the discussion in this section on the case
of planar Couette flow, returning to consider planar Poiseuille flow
later in the manuscript. We also focus on the slow flow limit $\Ca\to
0$, returning later on to consider the effects of finite $\Ca$.

\begin{figure}
  \centering
  \centerline{\includegraphics[scale=0.8,angle=0]{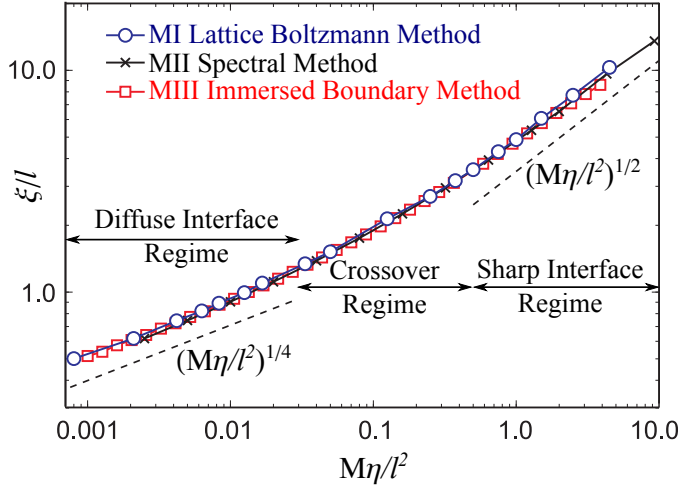}}
  \caption{Data for the slip length $\xi$ from all three numerical
    methods for planar Couette flow clearly demonstrating both the
    diffuse- and sharp-interface limits for small and large
    $M\eta/l^2$ respectively. All the results here are in the slow
    flow regime, $Ca < 0.01$, where the slip length is independent of
    the shear velocity. }
  \label{fig:Slip}
\end{figure}

The results are shown in Fig.~\ref{fig:Slip}. As can be seen, the
three numerical techniques give virtually indistinguishable results
across all four decades of $M\eta/l^2$. Over these four decades we can
distinguish two distinct regimes: a (i) diffuse interface regime when
$M\eta/l^2 \ll 1$, in which $\xi \propto (l_Dl)^{1/2}$, and (ii) a
sharp interface regime when $M\eta/l^2 \gg 1$, in which $\xi\propto
l_D$. In between these two distinct regimes is a broad crossover
window that itself spans around the decade either side of
$M\eta/l^2\approx 0.2$. In this way, importantly, our results
encompass and unify both of the two previously apparently
contradictory scalings put forward in the literature, by showing each
to apply in a different regime of the relevant dimensionless control
parameter $M\eta/l^2$.

In addition to the two basic lengthscales $l_D$ and $l$ present in the
model equations, out of which the slip lengthscale $\xi$ emerges in
the manner just described, there are two other lengthscales present in
our simulations, set by the system size $L_y$ and the discretisation
scale $\Delta{}x$.  We have ensured that the results presented in
Fig.~\ref{fig:Slip} are independent of possible finite size effects
due to $L_y$ and discretisation errors due to $\Delta{}x$.  The former
is a particular hazard in the limit of large $M\eta/l^2$, where the
slip length $\xi$ becomes large, and the latter in the limit of small
$M\eta/l^2$, where the slip length becomes small.

Having shown our numerical results to capture both the sharp interface
limit of $\xi\propto l_D$ for large $M\eta/l^2$, and the diffusive
interface limit of $\xi \propto (l_Dl)^{1/2}$ for small $M\eta/l^2$,
we are now in a position to reprise carefully the scaling arguments
put forward in the earlier literature for each of these two scaling
forms separately, and to discuss the validity of the assumptions made
in arriving at these functional forms.

We will follow a similar line of arguments as in \cite{Yue2010}, where
the Stokes and steady-state Cahn-Hilliard equations are integrated in
the $x$-direction along the wall. For concreteness, let us focus on the contact line
at the top wall, i.e. at $y = L_y$. The same scaling analysis applies for the
bottom wall ($y = 0$). For the $x$-component of the Stokes equation, this gives
\begin{equation}
  \int_{-\infty}^{\infty} \left[\eta \frac{\partial^2 v_x}{\partial y^2} + \eta \frac{\partial^2 v_x}{\partial x^2} - \frac{\partial p}{\partial x} - \phi \frac{\partial \mu}{\partial x} \right] dx = 0. \label{eq:sharpNS}
\end{equation}
The second term can be integrated to $\eta \partial v_x / \partial x |_{-\infty}^{\infty} = 0$.
The contribution of the third term in the integral is also zero, because
the pressure attains the same, constant value on either side of the interface, far from the contact line.
The fourth (final) term scales as $\mu_{\rm{max}}$, which represents the magnitude
of the chemical potential for a given flow setup.

The first term in Eq. \ref{eq:sharpNS} needs to be analysed more carefully.
In the sharp interface limit, the term $\partial^2 v_x/\partial y^2$
scales as $V_0/\xi^2$, and this term is significant across a length scale
given by the slip length $\xi$. Thus, we have
\begin{eqnarray}
  & \eta \frac{V_0}{\xi^2} \xi \sim \mu_{\rm{max}}, \nonumber \\
  & \mu_{\rm{max}} \sim \eta V_0/\xi. \label{eq:sharp1}
\end{eqnarray}
In the diffuse interface limit, the fluid velocity still varies across a length scale
$\xi$ in the $y$-direction, and as such, the term $\partial^2 v_x/\partial y^2$
still scales as $V_0/\xi^2$. However, to capture the key physics, the spatial
window of integration  must be broadened to the interface width $l$, since here
$\xi$ is {\em less} than the interfacial width $l$: recall the bottom leftmost data
points in Fig.~\ref{fig:Slip}. Taking the suitable of window of integration into account,
Eq. \ref{eq:sharpNS} leads to
\begin{eqnarray}
  & \eta \frac{V_0}{\xi^2} l \sim \mu_{\rm{max}}, \nonumber \\
  & \mu_{\rm{max}} \sim \eta V_0 l/\xi^2 . \label{eq:diffuse1}
\end{eqnarray}

Let us now carry out a similar analysis for the Cahn-Hilliard equation,
\begin{equation}
  \int_{-\infty}^{\infty} \left[ v_x \frac{\partial \phi}{\partial x} + v_y \frac{\partial \phi}{\partial y} - M \frac{\partial^2 \mu}{\partial x^2} - M \frac{\partial^2 \mu}{\partial y^2}\right] dx = 0. \label{eq:sharpCH}
\end{equation}
The first term can be integrated to give $V_0 \,  \phi|_{-\infty}^{\infty}  = 2 V_0$.
The second term is zero simply because $v_y = 0$ at $y = 0$. The third term integrates to
$-M\partial\mu / \partial x|_{-\infty}^{\infty}$, which is again zero because $\partial\mu / \partial x = 0$
far from the interface.

The result for the fourth term depends on the slip length regimes. $\partial^2 \mu / \partial y^2$ scales
as $\mu_{\rm{max}}/\xi^2$. However, the window of integration is different in the sharp ($\xi$) and diffuse
($l$) interface limits. For the sharp interface limit, the Cahn-Hilliard equation leads to the scaling
\begin{equation}
  V_0 \sim M \mu_{\rm{max}}/\xi. \label{eq:sharp2}
\end{equation}
On the other hand, for the diffuse interface limit, we obtain
\begin{equation}
  V_0 \sim M \mu_{\rm{max}}l/\xi^2 . \label{eq:diffuse2}
\end{equation}

We are now in the position to derive the scaling law predictions for the slip length.
Combining Eqs. \eqref{eq:sharp1} and \eqref{eq:sharp2} gives
\begin{equation}
  \xi \sim (M\eta)^{1/2} = l_D\label{eq:sharpscaling}
\end{equation}
in the sharp interface regime. Similarly, combining Eqs. \eqref{eq:diffuse1} and
\eqref{eq:diffuse2} for the diffuse interface regime results in
\begin{equation}
  \xi \sim (M\eta l^2)^{1/4} = (l_D l)^{1/2}.\label{eq:diffusescaling}
\end{equation}

As seen in Fig.~\ref{fig:Slip}, in between the sharp and diffuse
interface regimes is a smooth crossover where the behaviour varies
smoothly from one scaling law to another. We define the crossover
point by fitting the power laws in the sharp and diffuse interface
regimes separately, and finding where these two fits intersect each
other. This occurs at $l_D/l = 0.38$, in broad agreement with the
sharp interface criterion $l_D/l > 0.25$ proposed by \cite{Yue2010}

We note finally that the values of the slip length presented in
Fig. \ref{fig:Slip} are independent of the wall velocity, and
correspondingly of the Capillary number $\Ca = \eta
V_0/\sigma$, provided $\Ca \lesssim 0.01$.  (Data not
shown.) We shall demonstrate this independence explicitly below for
the case of planar Poiseuille flow (see Fig.~\ref{fig:Poiseuille}).

\subsection{Influence of a curvature term}
\label{sec:curvature}

In this section we test the extent to which our results for the slip
length do or do not depend on the microscopic details of the diffuse
interface model. Intuitively we might expect the slip length to be
independent of microscopics in the sharp interface regime, but
dependent on microscopics in the diffuse interface regime. Our
numerical results below provide evidence in favour of this intuition,
to which we shall however also add a note of caution at the end of
this section.

\begin{figure}
  \centering
  \centerline{\includegraphics[scale=0.7,angle=0]{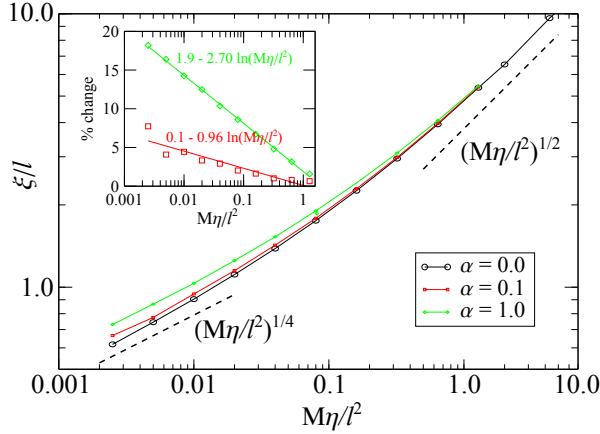}}
  \caption{Numerical data demonstrating how slip lengths in the
    diffuse-interface limit are affected by the inclusion of a
    curvature free energy term. The results of the standard Landau
    theory ($\alpha = 0$) are shown for comparison. Inset: percentage
    change relative to the case without curvature for $\alpha = 0.1$
    (red squares), $\alpha = 1$ (green circles). These data are
    obtained using the spectral method (MII) in planar Couette flow.}
  \label{fig:curvature}
\end{figure}

There are many possible ways to modify the basic Landau theory for which we presented results in the previous
subsection. We focus here on the simple but non-trivial extension made
by introducing the additional interfacial curvature term with a
strength set by $\alpha$ in Eq. \eqref{eq:fecurvature}.  Fig.
\ref{fig:curvature} shows our numerical results for $\alpha = 0.1$ and
$1.0$. Shown in the same plot for comparison are our original results,
already presented in the previous subsection, for the original theory
with $\alpha = 0$. As can be seen, the introduction of the curvature
term affects the slip length quite strongly in the diffuse interface
regime. This is to be expected: in this regime the physics of the
problem is determined not only by macroscopic quantities, but by the
microscopic gradient contributions to the free energy. In contrast, as
the control parameter $M\eta/l^2$ increases into the sharp interface
regime the dependence of the slip length on this microscopic parameter
$\alpha$ dramatically decreases and appears to become negligible in
the sharp interface limit $M\eta/l^2\to\infty$.

Although these numerical results strongly suggest that the slip length
is independent of the microscopic (gradient) contributions to the free
energy in the sharp interface regime, we attach the following note of
caution. In the case of equilibrium phase coexistence in the absence
of flow, it is possible to show by integrating across the interface
between the phases that macroscopic bulk quantities, such as the value
of the chemical potential at phase coexistence, are independent of the
structure of that interface, as determined by the spatial gradient
terms in the free energy.  However the same is not true in general for
the case of phase coexistence under the non-equilibrium conditions of
an applied flow because the equations of motion are usually
non-integrable in that case: they cannot be integrated across
  the interface to give a result that depends only on the bulk
  quantities on either side of the interface, but the result instead
  retains a dependence on the structure of the interface itself.  This
issue has been discussed in particular detail in the case of stress
selection in shear banded flows~\citep{Olmsted1999,Lu2000}.
Therefore, while our numerical results do strongly suggest that the
slip length is independent of the gradient terms in the sharp
interface limit, this question merits some further attention in future
studies.

\subsection{Relation to Cox's result}
\label{sec:Cox}

The results presented so far have demonstrated that, in the sharp
interface limit, the emergent slip length is determined by the
macroscopic dynamical variables $\eta$ and $M$.  While this is clearly
an important finding in terms of our physical understanding of the
moving contact line problem, in numerical practice this sharp
interface limit $M\eta/l^2 \rightarrow \infty$ can be difficult to
attain. Indeed the time taken to attain a steady state after the
inception of flow increases dramatically with increasing $M\eta/l^2$.
Lengthy simulations were required to obtain the rightmost 
data points in Fig.~\ref{fig:Slip}, with the extreme cases requiring run 
times of the order of weeks with a parallelised code using 4 cores of
Intel(R) Xeon(R) 2.40Ghz.


With this in mind, we now turn to address two questions of practical
numerical importance: (i) whether simulations carried out in the diffuse
interface regime still reproduce the expected macroscopic dynamics far
from the contact line, and (ii) whether our definition of the slip length is
still meaningful outside the sharp interface regime. We shall address
these questions in the context
of the Poiseuille flow protocol sketched in Fig.~\ref{fig:schematic}b,
in particular asking whether the dependence of the observed dynamic
contact angle on flow velocity is the same in the diffuse interface
regime as in the sharp interface regime.

\begin{figure}
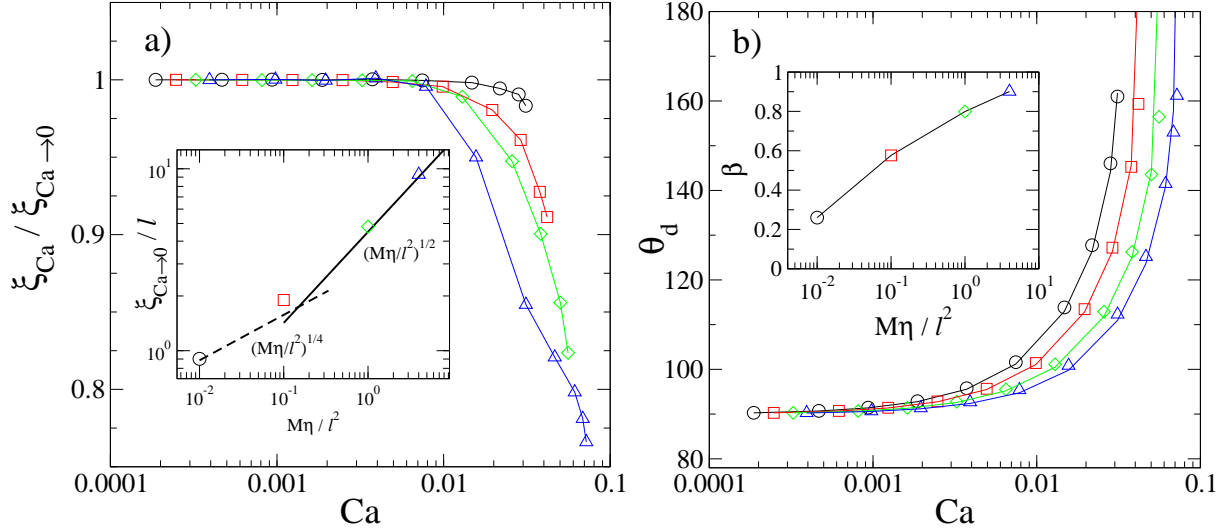

  \includegraphics[width=0.5\textwidth]{slip_4.eps}
  \includegraphics[width=0.47\textwidth]{fit_4.eps}
  \caption{Numerical data for the Poiseuille flow protocol, for $M\eta / l^2 = 0.01$ (black circles $\bigcirc$), $0.1$ (red squares), $1$ (green diamonds), $4$ (blue triangles). a) Demonstration that the slip length converges to a constant value as $\Ca \to 0$. The inset shows that the measured value of $\xi_{\Ca \to 0}$ using the Poiseuille flow protocol is in good agreement with the best fit power laws obtained for shear flow in Fig~\ref{fig:Slip}. b) Symbols mark the dynamic contact angle $\theta_d$ as measured from simulation. The solid lines show the analytical predictions of \cite{Cox1986} where $\delta = \beta \xi_{\Ca} / L_y$. The inset shows the best fit values of $\beta$ which appear to converge $\beta \to 1$ in the sharp interface limit, $M\eta/l^2 \to \infty$. The simulation data are obtained using the spectral method (MII). }
  \label{fig:Poiseuille}
\end{figure}

We start by checking that the slip length obtained for the Poiseuille
flow protocol is equivalent to that for planar Couette flow.  Fig.
~\ref{fig:Poiseuille}a shows that the slip length for the Poiseuille
case converges to a constant value as $\Ca \to 0$, as for the planar
Couette case.
We have checked in this regime that the measured values of the slip
length for the Couette and Poiseuille flows are within 0.5\% of each
other over the whole range of $M\eta/l^2$. See the inset of
Fig.~\ref{fig:Poiseuille}a, where the slip lengths for the Poiseuille
flow are compared to the best fit power laws extracted for planar
Couette flow in both the diffuse and sharp interface regimes. For
larger $\Ca > 0.01$ the slip length starts to depend on the fluid
velocity, as in planar Couette flow.

We now turn to the dynamic contact angle, as defined in
Fig.~\ref{fig:schematic}b.  It is known that this angle increases with
increasing flow speed: for increasing pressure drop $\Delta P$ in
Fig.~\ref{fig:schematic}b the interface is deformed more significantly
due to the larger flow-speed mid channel compared to the stationary
wall, and correspondingly the dynamic contact angle deviates more from
the equilibrium value.  Associated with this is a larger gradient in
the chemical potential, which indeed allows the contact line to move
at a higher velocity.

An analytical prediction for this increase of the dynamic
contact angle with flow speed, as characterised by the capillary
number $\Ca$, was given by~\cite{Cox1986}:
\begin{equation} \label{eq:Cox}
  g(\theta_d) = g(\theta_e) + \Ca \ln(\delta^{-1}).
\end{equation}
Here $\theta_d$ is the dynamic contact angle and $\theta_e$ is the
contact angle at equilibrium (here $90^\circ$). For fluids of
matched viscosity the function $g$ is defined as
\begin{equation}
  g(\theta) = \int_0^\theta \frac{ \pi\phi(\pi-\phi) + (2\pi\phi-\pi^2)\sin{\phi}\cos{\phi} - \pi \sin^2{\phi} } { 2\pi^2\sin{\phi} } \,\, d\phi.
\end{equation}
The parameter $\delta$ in Eqn.~\ref{eq:Cox} is the ratio of the
microscopic slip length to some characteristic macroscopic lengthscale
of the system. We choose that to be the channel width $L_y$ and set
$\delta = \beta \xi_{\Ca} / L_y$ where $\beta$ is a constant fitting
parameter.

To test Cox's prediction, we plot our simulation results for the
dependence $\theta_d$ on the capillary number $\Ca$ in
Fig.~\ref{fig:Poiseuille}b. We do so for four different values of
$M\eta/l^2$, which we ensured span both the diffuse and sharp
interface limits, as well as the crossover regime between them.
Pleasingly we find excellent agreement with Cox's prediction across
the the full range of $\Ca$, for all values of $M\eta/l^2$.
Furthermore, as shown in the inset of Fig.~\ref{fig:Poiseuille}b, the
fitting parameter $\beta$ appears suggestively to converge to unity in
the sharp interface limit, $M\eta/l^2 \to \infty$.

These results suggest that simulations in the diffuse interface limit
(and similarly the crossover regime) still provide a reliable
representation of the macroscopic dynamics. However, the effective
slip length must be suitably corrected by an $O(1)$ prefactor $\beta$.
Increasingly minor corrections are needed as one approaches the sharp interface limit,
and we expect Cox's result would be directly obtained for $M\eta/l^2
\to \infty$, where $\beta\to 1$.

\section{Conclusions}
\label{sec:conclusion}

In this paper, we have performed extensive numerical simulations to
reconcile previously apparently contradictory scaling laws for the
slip length associated with a moving contact line in diffuse interface
models.  We did so by exploring four decades of the relevant
dimensionless control parameter $M\eta/l^2 = l_D^2 /l^2$, and three
decades of the capillary number $Ca$. Moreover we used three
independent numerical methods, demonstrating their predictions to be
in excellent quantitative agreement across all decades.

In the limit of slow flows corresponding to typical capillary numbers
$Ca < 0.01$, we find that the slip length converges to a well-defined
value that is independent of the flow speed. In this slow flow regime
we explored four decades in $M\eta/l^2 = l_D^2 /l^2$.  Doing so
enabled us to distinguish two distinct regimes for the scaling of the
slip length, and a broad crossover window in between. For $M\eta/l^2
\ll 1$ we found a diffuse interface regime in which the slip length
scales as $\xi \sim (l_Dl)^{1/2}$, consistent with the previous result
of \cite{Briant2004b}, and is furthermore sensitive to form of the
microscopic interfacial terms in the free energy functional. In
contrast for $M\eta/l^2 \gg 1$ we recover the sharp interface limit
previously discussed by \cite{Jacqmin2000} and \cite{Yue2010}. Here
the slip length is proportional to the diffusion length scale,
$\xi\propto l_D \sim (M\eta)^{1/2}$, depending only on the macroscopic
dynamical variables of viscosity $\eta$ and molecular mobility $M$.
Our numerical results further suggest the slip length to be
insensitive to changes in the microscopic interfacial free energy
terms in this sharp interface regime.

For $Ca > 0.01$, the effective slip length decreases with increasing
capillary number, in qualitative agreement with the work of  \cite{Chen2000}.
However, we are not able to reproduce their prediction $\xi\propto V_0^{-1/2}$
of in this fast flowing regime. More broadly, so far we are unable to
find any reliable scaling law that is valid across a broad range of Capillary number,
and for both the sharp and diffuse interface regimes.

Our numerical results also allow us to appraise suitable sets of
simulation parameters.  By comparing our simulation results for the
dynamic contact angle to the analytical prediction by \cite{Cox1986},
we find excellent agreement if we allow the slip length to be rescaled
by a dimensionless prefactor $\beta=O(1)$, which pleasingly appears to
converge to unity in the sharp interface limit. In numerical practice,
however, the sharp interface limit can be very expensive to realise,
partly because the time taken to attain a steady state increases
dramatically with increasing $M\eta/l^2$, and partly because the slip
length is O(10) larger than the interface width, with demanding
implications on the overall size of the simulation box.  In the
diffuse interface limit, in contrast, the timescale and lengthscale
requirements are less stringent, and simulations accordingly much less
expensive to perform. Our fits to the Cox formula suggest that a
reliable representation of the macroscopic physics is nonetheless
attainable even in the diffuse regime, provided the lengths are
suitable rescaled by this corrective prefactor $\beta=O(1)$.

There are several avenues for future work.
Firstly, it would be interesting to investigate how
the two scaling regimes for contact line slip in diffuse interface
models affect more complex flow problems, such as
the macroscopic dynamics of a liquid droplet
\citep{Servantie2008,Mognetti2010,Thampi2013} or the criterion for
the liquid entrainment (wetting failure) \citep{Eggers2004,
Sbragaglia2008}, and whether similar renormalization of the slip
length is adequate to account for their differences, as for the
Poiseuille flow considered here. In these cases, we need to
implement non-neutral wetting boundary conditions.
We expect the prefactors of the scaling laws to change with contact angles,
but not the exponents themselves.

Secondly, given the crossover we have reported in this study,
it would be insightful to revisit the liquid-gas system and understand
the sharp interface limit of this model.
In the binary model the slip mechanism is diffusion, and as such,
the slip length depends on the mobility parameter, in addition to the
fluid viscosity and the interface width. In the liquid gas system, the mechanism
is evaporation-condensation instead, and we expect the slip length to
depend on the density ratio between the liquid and gas phases.

Thirdly, we note that we have taken the zero temperature limit in this work and
have not accounted for thermal fluctuations. It will be interesting to study if such fluctuations
simply broaden the effective interface width, or they give rise to new phenomena.
For example, enhanced spreading due to thermal fluctuations has been reported by \cite{Davidovitch2005} and \cite{Gross2013}.

Finally, experimental systems
consisting of colloid-polymer mixtures are known to phase separate
into colloid-rich and colloid-poor domains, with an interface width that is
of order 1 micron \citep{Aarts2004,Ledesma2015}, much larger
than the typical value for molecular fluids. Such a system can potentially
be exploited to realise the different contact line slip regimes discussed here experimentally.

Acknowledgements: The research leading to these results has received funding from the European
Research Council under the European Union's Seventh Framework
Programme (FP7/2007-2013) / ERC grant agreement number 279365.

\bibliographystyle{rsc}
\bibliography{refs}

\end{document}